\documentstyle[amssymb,pre,aps]{revtex}
\begin{document}
\twocolumn[\hsize\textwidth\columnwidth\hsize\csname@twocolumnfalse\endcsname

\title{Deterministic growth model of Laplacian charged particle aggregates}
\author{
N.I. Lebovka$^{1,2}$\thanks{lebovka@roller.ukma.kiev.ua},Y.V. Ivanenko$^1$,
N.V. Vygornitskii$^1$}
\address{$^1$Institute of Biocolloidal Chemistry, NAS of Ukraine, 42, blvr.\\
Vernadskogo, Kyiv, 252142, Ukraine\\
$^2$Kyiv Mogyla Academy University, 2, vul. Scovorody, Kyiv, 252145, Ukraine}
\draft
\date{Received \today}
\maketitle

\begin{abstract}
The results of the computer simulation of the aggregates growth of the
similarly charged particles in the framework of deterministic Laplacian
growth model on a square lattice are presented. Cluster growth is controlled
by three parameters ${p, E,\lambda}$, where $p$ - Laplacian growth parameter,
$E$ - energy of a particle sticking to a cluster, $\lambda$ - the screening
length of electrostatic interactions. The phase diagram of cluster growth is
built in the co-ordinates ${E,\lambda}$. The zones of different cluster
morphology are selected: I-the zone of finite X-like structures,II-the zone
of infinite ramified structures, controlled by electrostatic interactions,
III-the zone of infinite structures with electrostatic interactions
effectively switched off. Simple electrostatic estimations of the locations
of the zone boundaries are presented. It is shown that in general case
within the zone II the continuous change of $D_f$, controlled by parameters 
${p, E,\lambda}$, takes place. In the degeneration limit when the given model
transforms into deterministic version of the Eden model (at $p=0$), the
crossover from linear $(D_f=1)$ to compact $(D_f=2)$ structures is observed
when passing through the boundary between the zones I and II.
\end{abstract}

\vskip 2pc]

% body of paper here
A number of models have been developed to describe the growth of aggregates
with complex, ramified structure. The most important among them are the
models of diffusion-limited aggregation fractal growth (DLA model), compact
growth (Eden model), ballistic growth, dendrite, and dense-branching
structure growth \cite{meakin}. On changing the local conditions controlling
the process of particle's sticking to a cluster, it is possible to observe
the structures with different morphologies \cite{erlebacher,ziquin,john}.
The study of aggregation, which takes into account the specific interactions
between the particles, e.g., the interactions of electric or magnetic
origin, is of a special interest. Such interactions may effectively manifest
themselves in colloids \cite{heglessen,fernandez,bos}, in electrochemical
deposition phenomena \cite{john}, in discharge phenomena (in particular, in
ball lightning development from unipolar plasma \cite{smirnov}).

This work presents the results of the growth study of aggregates formed by
the likely charged particles within the framework of deterministic Laplacian
growth model. Such model is widely used to describe the growth of
dendrite-like structures \cite{family,vygornitskii}. The model of
such type allows to obtain the deterministic analogue of DLA-type fractal
patterns. The choice of deterministic model allows also to enhance the
lattice-imposed anisotropy of the patterns and to suppress the noise effects.

Our model takes into account the screening effects by limiting the radius of
electrostatic repulsion between the particles as $r<\lambda $. Each
simulation step included the numerical solving of Laplace equation on a
lattice and the calculation of sticking probability $f$ ($\propto $ to the
normalized gradient of the potential) all over the cluster perimeter. Before
the sticking of a new particle to the point $(j)$ of a cluster, the
calculation of electrostatic repulsion energy $U(j)$ between the new
particle and all the other particles of cluster formed on the previous
evolution step was carried out:

\begin{equation}
U(j)=\sum\limits_i{q^2a_{ij}/r_{ij}},  \label{eq1}
\end{equation}
where $q$ is a particle charge, $r_{ij}$ is a distance between the $i$
particle of the aggregate and the $j$-point, $a_{ij}=1$ at $r_{ij}$ $\lambda 
$ and $a_{ij}=0$ otherwise.

New particle joined the cluster at the $(j)$-point of perimeter only when
two conditions were satisfied: a) $f>p$ in this point, and b) $U(j)$ did not
exceed some critical value $E$. Here, $p$ is the Laplacian growth parameter, 
$0<p<1$. Note that $E$ corresponds to the short-range energy of single
particle attraction to a cluster, and that this attraction may result from
the Van der Waals interactions.

The growth of clusters took place on the two-dimensional square lattice. The
distances were measured in the lattice units, and all the particles had a
charge $q$ equal to 1. Thus, we may consider that $\lambda ,E$ and $U$
parameters introduced above are dimensionless. The lattice size comprised
1000$\times $1000 and the number of cluster particles did not exceed $5\cdot
10^5$.

So, in this model the three parameters,$\lambda ,E,p$ control the growth of
clusters. In the regular deterministic Laplacian growth model, assuming no
electrostatics, the growth of clusters is controlled only by the value of
parameter $p$ and their size is not limited. At $p=0$, our model corresponds
to deterministic Eden model and gives the compact clusters with fractal
dimensionality $D_{fo}=2$. At $0<p<1$, the structures with fractal
dimensionality develop; their dimensionality decreases with the increase of
p \cite{family},\cite{vygornitskii}. For $p=0.3$, which case will be
analyzed in more details below as an example, we have obtained, $D_{fo}=1.49$%
.

The switch-on of electrostatic interactions results in considerable
complication of the growth patterns. In our case, both finite and infinite
clusters may form, depending on the values of ${\lambda ,}E,p$ parameters.

Figure \ref{fig1} presents the phase diagram in $\lambda -E$ coordinates,
where the zones of different cluster morphology are distinguished. The
boundary between the zones II and III is shown by the line $1$ for $p=0.3$
and by the line $1^{\prime }$ for $p=0$. The cluster patterns from the
different zones of diagram (marked by the symbol $\blacksquare $ and
relevant Latin characters) are presented on Fig.\ref{fig2}. All clusters
have specific anisotropic square-like shape due to deterministic version of
a growth model for clusters growing on a square lattice and in a given
presentation of clusters the lattice axes are directed along the square
diagonals.

In the zone I, the growth of finite X-like structures is observed, and in
the zones II and III the growth of infinite clusters takes place. The
maximal cluster size limitation observed in the zone I may be explained as
follows: at high enough $\lambda $ values (or, correspondingly, small enough 
$E$) the electrostatic repulsion energy at the points of new particles
sticking $U$ continuously grows with cluster growth. At some critical value
of cluster size the realization of condition $U>E$ occurs at all the sites
of cluster perimeter, which means the halt of the cluster growth. The
development of non-ramified X-like structures in the zone I may be explained
by almost complete control of the cluster growth by electrostatic repulsion
forces in the given area of $\lambda ,E$ values. Under such conditions, the
particles stick to the cluster predominantly at points most distant from the
cluster center, where the energy of electrostatic repulsion is the lowest.
In this case, the growth of clusters proceeds mainly along the lattice axes,
i.e., on the lines directed along the diagonals of the square.

At $\lambda =\infty $, only the growth of finite clusters with the size
controlled by parameters $E$ and $p$ takes place, and the ramification of
structures is also possible. The dash-filled area on Fig.\ref{fig1}
approximately corresponds to the zone of transition to the areas of infinite
cluster growth. Note that the location of this zone practically does not
depend on the value of the parameter $p$.

The zone of infinite clusters, where fractal, dense-branching and dense
structures develop, should be divided into two areas - II and III. The
necessity of such division follows from the detailed analysis of fractal and
morphological properties of clusters in these areas.

The estimation of the fractal dimensionality $D_f$ of the deterministic
growth patterns requires some caution. We used the ''sand-box'' method for $%
D_f$ estimation and analyzed the number of particles in the squares of
ascending size \cite{pfeifer}. Application of the squares instead of circles
commonly used in this method allows to consider more precisely the effects
of cluster anisotropy imposed by the symmetry of square lattice. Besides, as
it is seen from Fig.\ref{fig2}, the pattern structures may be subdivided
quite clearly, especially at large $E$ values, into the dense X- like
nucleus (with $D_f=2.0 $) and ramified peripheral part. We calculated the
fractal dimensionality only for ramified part of a pattern.

Figure \ref{fig3} presents dependencies of $D_f$ on $E$ for $\lambda =10,100$
(curves $1$ and $2$) and $D_f$ on $\lambda $ for $E=100$ (curve $3$) in the
case of $p=0.3$. When moving across the phase diagram (Fig.\ref{fig1}) along
the $E$ axis at some constant value of $\lambda $ (same as at $\lambda
=\infty $), the clusters with fractal dimensionality $D_f=1$ are observed in
the zone I. The $D_f$ sharply increases on passing into the zone II; further
on,the decreasing of $D_f$ takes place and at some point $E=E_{II-III}$,
which corresponds to the point of transition into zone III, the fractal
dimensionality reaches the value $D_{fo}=1.49$. As it was mentioned above,
this value corresponds to the fractal dimensionality of clusters that grow
at $p=0.3$ without electrostatic interactions. When moving along the $%
\lambda $ axis at some fixed value of $E$, the clusters characterized by the
constant value of fractal dimensionality, $D_{fo}=1.49$ are found in the
zone III. But at some point $\lambda =\lambda _{II-III}$, which corresponds
to the point of transition into the zone II, a rather sharp increase of $%
D_{fo}$ takes place.

The similar behavior of the fractal dimensionality is observed for other
non-zero values of $p$, though localization of the boundary between the
zones II and III depends on the $p$ value. The $D_f$ behavior somewhat
differs at $p=0$. In this case, the clusters of the zone I are characterized
by $D_f=1$ (same as for $\lambda =\infty $), but $D_f=2$ in the zones II and
III. Thus, on passing from the zone I into the zone II across the
dash-filled zone, a crossover between the linear ($D_f=1$) and dense
structures ($D_f=2$) is observed. The clusters of the zone II have a
dense-branching morphology and their structure becomes more compact as far
as the system approaches the boundary of the zone III (the curve $1^{\prime
} $ on the Fig.\ref{fig1}). Zone III is characterized by the growth of
compact clusters of the square shape only, which is practically equivalent
to the cluster growth conditions with switched-off electrostatic
interactions.

As it follows from the above discussion, zone II of the phase diagram
corresponds to the strong effect of electrostatic interactions on a cluster
growth, whereas zone III is the zone of cluster growth under completely
switched-off electrostatic interactions. Here, we propose the simple
electrostatic estimations for explanation of the observed behavior.

The absence of any considerable effect of electrostatic interactions on the
growth of clusters inside the zone III may occur only if the points of
cluster-growth halt, complying with the condition of $U>E$, are absent on
the cluster perimeter. Then, the boundary between the zones II and III may
be estimated out of the condition $E_{II-III}<U$. The highest electrostatic
repulsion is peculiar to perimeter points coinciding with the middle points
of the square sides (Fig.\ref{fig2}), since these perimeter points are the
nearest to the cluster center. To calculate the electrostatic repulsion
energy $U$ in these points, we pass from summation to integration in
relation (\ref{eq1}) and assume the ideal density profile for fractal
clusters on receding from the center

\begin{equation}
\rho (R)\approx R^{D_f-2},  \label{eq2}
\end{equation}
where $R$ is the distance from the cluster center to the middle-point of the
cluster side.

The calculations show that

\begin{equation}
E_{II-III}\approx U=\pi \lambda ^{D_f-1}f(D_f),  \label{eq3}
\end{equation}
where $f(D_f)$ is the function of cluster fractal dimensionality, $D_f$ .
Figure \ref{fig4} demonstrates the form of this function $f(D_f),$ obtained
through the numerical integration.

Equation (\ref{eq3}) defines the theoretical boundary between the zones II
and III., This boundary is shown on the Fig.\ref{fig1} for the case of $p=0.3
$ by the dashed line $2$. We see that this theoretical boundary line goes
above the real boundary shown by the line $1$, which is not surprising since
the density profile assumption (\ref{eq2}) does not take into account the
existence of the compact nucleus.

On $p=0$, i.e., for the case of deterministic analogue of the Eden model,
the growth of compact clusters is observed, and we obtain from (\ref{eq3}%
)

\begin{equation}
E_{II-III}\approx U=\pi \lambda .  \label{eq4}
\end{equation}

Condition (\ref{eq4}) defines the boundary between the zones of compact
clusters growth and dense-branching morphology. In this case, the
theoretical boundary defined by (\ref{eq4}) perfectly describes the
simulation data (line $1^{\prime }$).

Authors thank Dr. N. S. Pivovarova for valuable discussions
of the manuscript  and help with its preparation.

% now the references. delete or change fake bibitem. delete next three
% lines and directly read in your .bbl file if you use bibtex.

% figures follow here

\begin{figure}[tbp]
\caption{Phase diagram of Laplacian charged particle aggregates for
deterministic growth model presented in the following co-ordinates:
screening length, $\lambda$ against sticking energy, $E$. The zones of
different pattern growth morphology are distinguished: I is the zone of
finite X-like structures, II is the zone of infinite ramified structures
with cluster growth controlled by electrostatic interactions, III is the
zone of infinite structures with effectively switched-off electrostatic
interactions. Symbol $\blacksquare$ marks the points corresponding to the
clusters presented on the Fig. \ref{fig2}, which were grown under the
following conditions: $\lambda=100 -E=14(a),40(b), 100(c)$; $\lambda=\infty
- E=50(a^{\prime}), 100(b^{\prime}), 500(c^{\prime})$. The solid lines 1 and
1' show the boundaries between the zones II and III for p=0.3 and p=0,
correspondingly. The dashed line 2 shows the theoretically predicted
position of this boundary for p=0.3.}
\label{fig1}
\end{figure}

\begin{figure}[tbp]
\caption{Typical cluster patterns for the model of deterministic Laplacian
growth of charged particles aggregates obtained at different values of $p,
E, \lambda$ (which are shown on the figure). Characters a, a', b, b', c and
c' correspond to the points, shown on the phase diagram (Fig.\ref{fig1}).}
\label{fig2}
\end{figure}

\begin{figure}[tbp]
\caption{Dependencies of the fractal dimensionality Df versus the sticking
energy $E$ at the fixed values of $\lambda=10(1), 100(2)$ and versus the
screening length $\lambda$ at the fixed value of $E$=10 (3).}
\label{fig3}
\end{figure}

\begin{figure}[tbp]
\caption{Profile of $f (D_f )$ from the equation (3). This function is used
for theoretical estimation of the boundary between the zones II and III on
the phase diagram (Fig.\ref{fig1}).}
\label{fig4}
\end{figure}

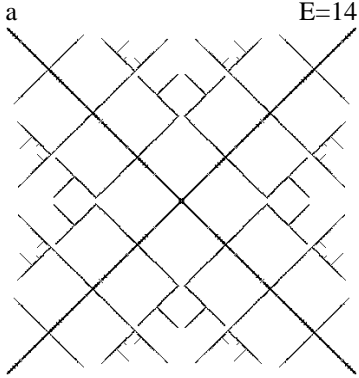
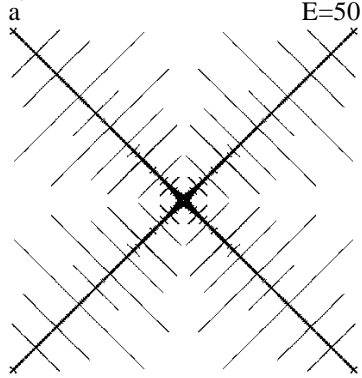
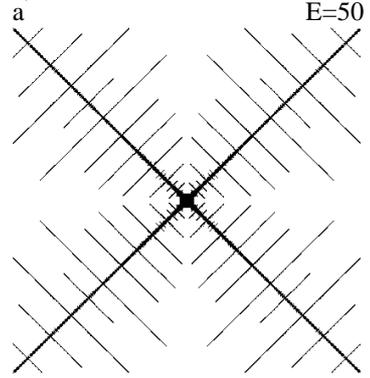
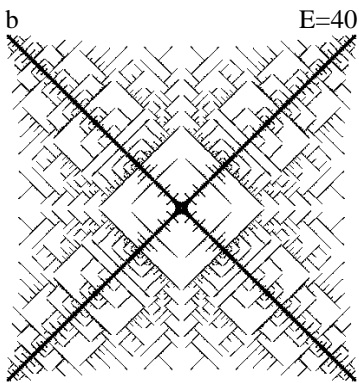
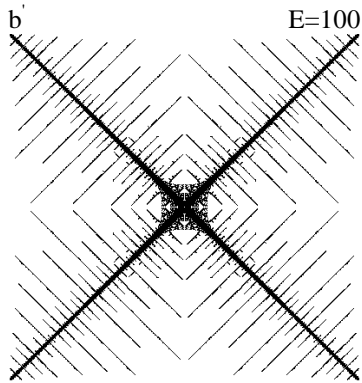
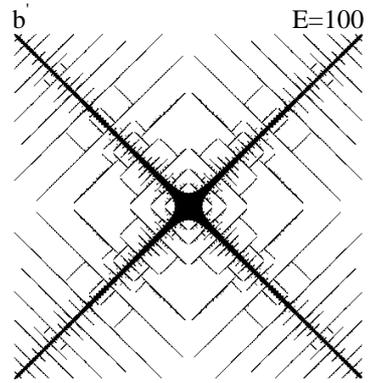
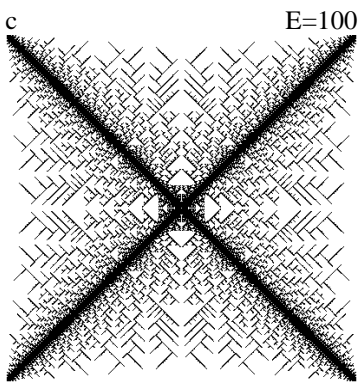
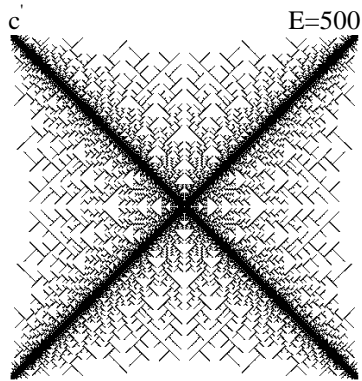
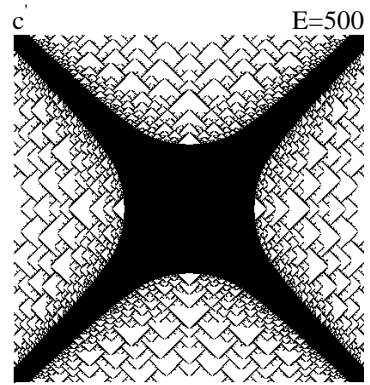
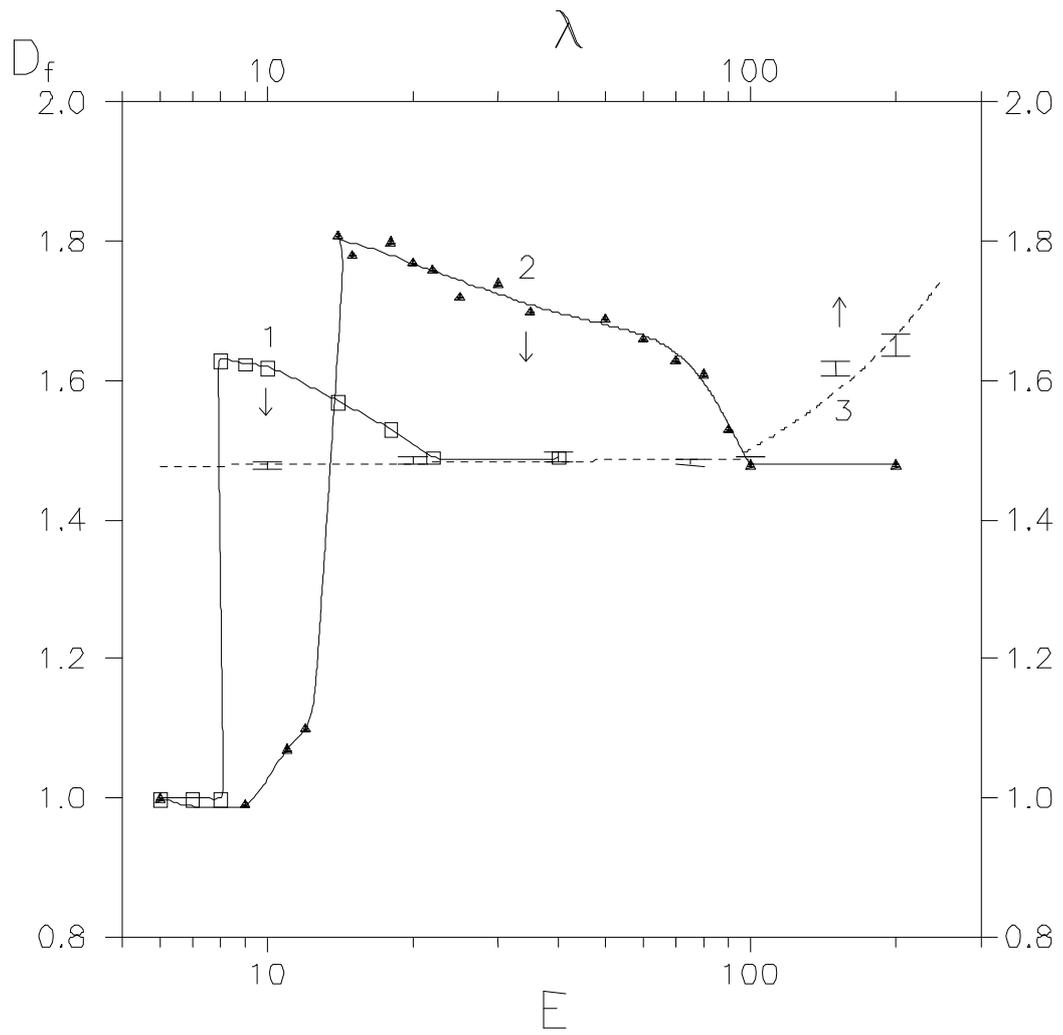
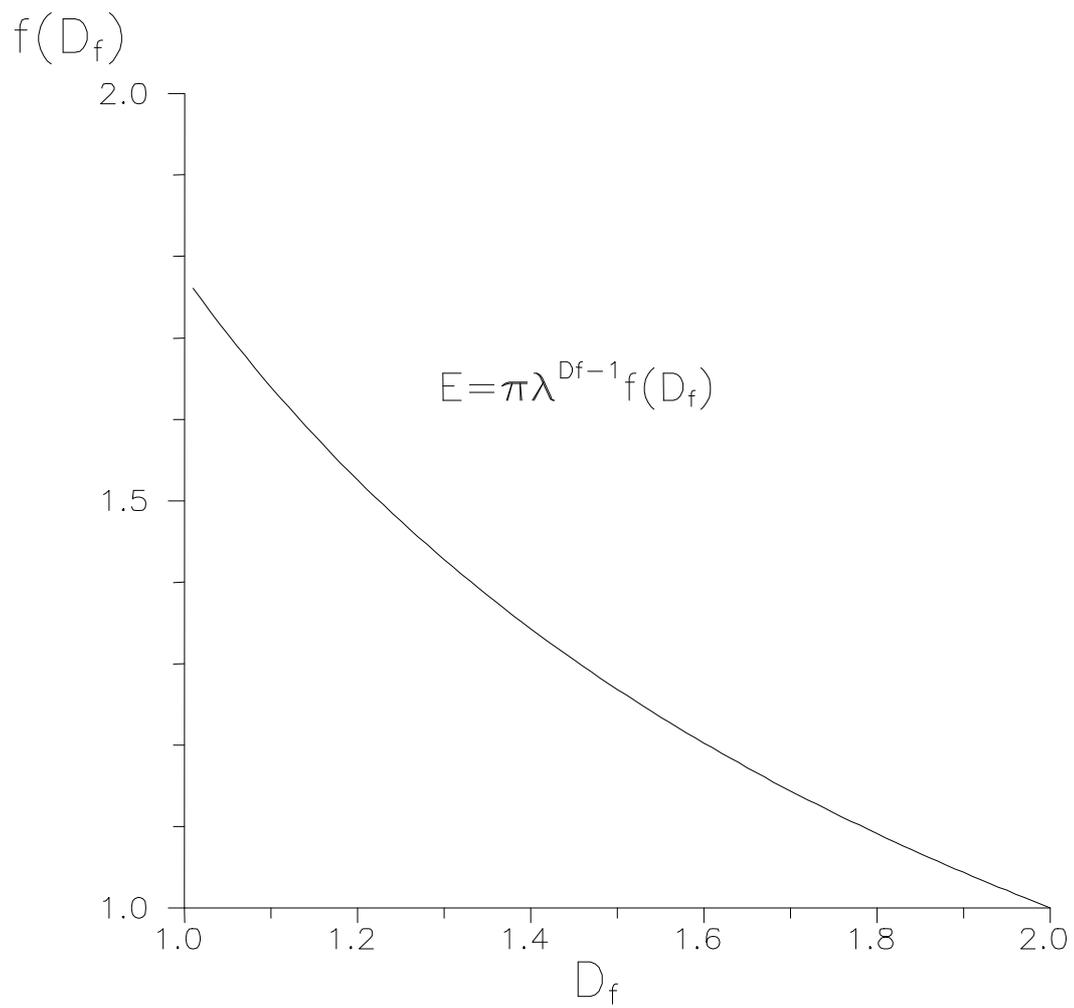
\end{document}